\documentclass{ws-procs10x7}
\usepackage{balance}

\newcolumntype{d}[1]{D{.}{.}{#1}}

\makeindex
\begin{document}

\title{Saturation effects in diffractive scattering at LHC energies }
\author{O. V. SELYUGIN}

\address{Bogoliubov Laboratory of Theoretical Physics, JINR, Dubna, RUSSIA
 \\E-mail: selugin@theor.jinr.ru}
\author{J.-R. CUDELL$^*$ }

\address{Physics Department, Universit\'e de Li\`ege, Li\`ege, BELGIUM
 \\$^*$E-mail: JR.Cudell@ulg.ac.be}


\twocolumn[\maketitle\abstract{Unitarization schemes can be reduced to
non-linear equations which
saturate at small $b$ the evolution of the elastic amplitude with $s$,
and which mimic parton saturation in the non-perturbative regime.
These equations enable us to study the effect of saturation on
total and elastic cross sections for various models, and to evaluate the
uncertainties on $\sigma_{tot}$ and $\rho(s,t)$ at the LHC in the presence of
a hard pomeron. }
\keywords{Elastic scattering; Saturation; Unitarization.}
]

\section{Introduction}

The most important results on the energy dependence of diffractive
hadronic scattering were obtained from first principles
(analyticity, unitarity and Lorentz invariance), which lead to
specific analytic forms for the scattering amplitude as a function
of its kinematical parameters - $s$, $t$, and $u$.
One of the important theorems is the Froissart-Martin bound
 which states that the high-energy cross section for the
scattering of hadrons is bounded by
 $ \sigma_{tot} \sim
 \log^2 (s/s_0) $,  
where $s_0$ is a scale factor.

Experimental data reveal that total cross sections
grow with energy. This means that the leading contribution in the
high-energy limit is given by the rightmost singularity, the pomeron,
with intercept exceeding unity. In the framework of perturbative QCD,
the intercept is also expected to exceed unity by an amount
proportional to $\alpha_s$
At leading-log $s$, one obtains
the leading singularity at $J-1 \  = \ 12 \ \log 2({\alpha_s }/{\pi})$.
In this case, the Froissart-Martin bound is soon violated.

  In a recent study,\cite{clms2,clms3} we have found that forward data,
including
 total cross sections and the ratios of the real part to the imaginary part of
  the scattering amplitude, are well fitted by a combination of a soft pomeron (which would be
  purely non perturbative) and a hard pomeron.  The inclusion of these two pomerons,
 together with the use of the integral dispersion relations,\cite{clms1} and the addition of
 the sub-leading meson trajectories, leads to a successful description of all
 $pp$, $\bar p p$, $\pi^\pm p$,
$K^\pm p$, $\gamma p$ and $\gamma\gamma$ data for $\sqrt{s}\leq 100$ GeV.
  Indeed, the fast growing hard pomeron leads to a violation of
 unitarity for values of $\sqrt{s}$ around 1 TeV.

   Now the saturation processes are actively studied in high-energy physics.
  They are connected with the saturation of the gluon density at small x,
which may be related to the Black Disk Limit (BDL).
  However, one should note that the process of saturation
has mostly been examined in the framework of the dipole model.\cite{muller}
  But, as noted by the Golec-Biernat and Wusthoff,\cite{CBWwud}
saturation in that case suppresses the soft contributions and the hard
Pomeron plays the main role. That is why
  the dipole model describes so successfuly hard processes
(such as vector-meson production).
  Contrarily, saturation in the BDL sense tames the hard Pomeron and increases
the role of the long-distance processes. It is the latter that we shall
study in this contribution.

The unitarity condition can be implemented via several different
prescriptions. Two of them are based on particular solutions of  the
unitarity equation.
The scattering amplitude in the impact parameter representation is defined as
\begin{eqnarray}
  T(s,t) = i \int_{0}^{\infty} \ b db J_{0}(b \Delta)
    f(b,s).
\end{eqnarray}
  with
 $   f(b,s) \ \leq \ 1.$
It satisfies the unitarity equation
\begin{eqnarray}
  Im f(s,b) &=& [Im (f(s,b)]^2 + [Re f(s,b)]^2 \nonumber \\
            & & + \ \ \ \eta_{inel}(s,b).
\end{eqnarray}
  One of the possibilities is obtained in the $U-$matrix approach
\cite{umat,trosh}:
\begin{eqnarray}
       f(s,b) \ = \ \frac{U(s,b)}{1\ - \ i \ U(s,b)}.
\label{U-matrix}
\end{eqnarray}

The second possible solution of the unitarity condition
corresponds to the eikonal representation
\begin{eqnarray}
  f(s,b) \ = \ [1 -  \exp(- \chi(s,b)].
\end{eqnarray}
One often takes the eikonal phase in factorised form
\begin{eqnarray}
 \chi(s,b) = h(s) \ f(b),
\end{eqnarray}
and one supposes that, despite the fact that the energy dependence of
$h(s)$ can be a power
  $ h(s) \sim s^{\Delta} $,
the total cross section will satisfy the Froissart bound
  $           \sigma_{tot} \leq \ a \ \log^2 (s) $
We find in fact that
the energy dependence of the imaginary part of the amplitude
and hence of the total
cross section will depend on the form of $f(b)$, {\it i.e.} on
the $t$ dependence of the slope of the elastic scattering
amplitude. If $f(b)$ decreases as
a power of $b$, the Froissart-Martin bound will always be violated.
In the case of other forms of the $b$ dependence,
a special analysis\cite{dif04} is required.

\section{ Non-linear equations}
The problem of the implementation of unitarity via saturation is
that the matching procedure seems arbitrary. Hence we considered
a different approach to saturate the amplitude.
It is connected with the non-linear saturation processes which have
been considered in a perturbative QCD context.\cite{grib,mcler}
Such processes lead to an infinite set of coupled
evolution equations in energy for the correlation functions of
multiple Wilson lines.\cite{balitsky}
In the approximation where the correlation functions for more than
two Wilson lines factorise, the problem reduces to the non-linear
Balitsky-Kovchegov (BK) equation.\cite{balitsky,kovchegov}

It is unclear how to extend these results to the non-perturbative region,
but one will probably obtain a similar equation. In fact we found simple
differential equations that reproduce either the U-matrix or the eikonal
representation.

We shall consider saturation equations of the general form\cite{cs7}
\renewcommand{\S}{{\cal S}}
\begin{equation}
{dN / d\xi}=\S(N)
\end{equation}
 We shall
impose the
unitarity conditions
 $N\rightarrow 1$ as $s\rightarrow\infty \ $ and
 $dN/d\xi\rightarrow 0$ as $s\rightarrow\infty. \ $
We shall also assume that $\S(N) \ $ has a Taylor expansion in $N$,
and that
the first term only the hard pomeron $N_{bare}=f(b) s^\Delta$,
where $\Delta$ is the pomeron intercept minus $1$.
Similarly, we fix the integration constant by demanding that the first
term of the expansion in $s^\Delta$ reduces to $N_{bare}$.
We then need to
take $\S(N)=\Delta N+O(N^2)$. The conditions at  $s\rightarrow\infty$ then give
$\S(N)=\Delta(N-N^2)$ as the simplest saturating function.
The resulting equation
\begin{equation}
{dN / d y}=\Delta (N-N^2),
\label{u-matr}
\end{equation}
 where $y=log(s) \sim log(1/x)$
This equation has the solution
\begin{equation}
N=\frac{f(b)s^\Delta}{f(b)s^\Delta + 1}
\end{equation}

One may then wonder whether other unitarization schemes are possible.
We give here a general algorithm to build such schemes:
\begin{equation}
\frac{dN}{ d y}=\frac{dN_{bare}}{d y}(1-N)
\end{equation}
with $N_{bare}$ the unsaturated amplitude.
This will trivially obey the conditions at  $s\rightarrow\infty$,
and saturate at $N=1$.
It has as a solution
\begin{equation}
N(b,s)=1-\exp(-N_{bare}(b,s))
\end{equation}
This is equivalent to the eikonal representation with
 $N_{bare}(b,s)$ corresponding the Born aplitude in the impact
 representation.
The corresponding equation can also be written in the form
\begin{equation}
\frac{dN}{d y}= \Delta \ N_{bare} \ \exp[-N_{bare}(b,s)].
\label{eik}
\end{equation}
  One can also find new unitarization schemes, {\it e.g.} from the equation
\begin{equation}
{dN / d y}= \frac{d N_{bare}}{dy} \ (1-N^2),
\label{tangh}
\end{equation}
one gets
\begin{equation}
N(b,s)= \tanh[N_{bare}(b,s)].
\end{equation}
   On Fig.1, the result of saturation is shown for these three
   unitarization procedures. It is clear that despite
the essentially difference in the forms of
   the non-linear equations, the process of the saturation is very similar.

\begin{figure}
\centerline{\psfig{file=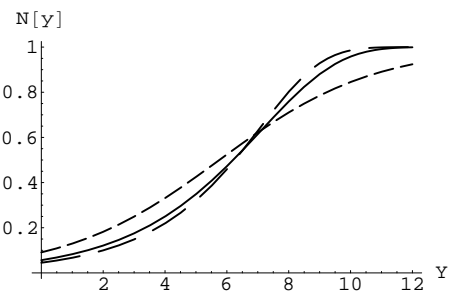,width=2.5in}}
\caption{The saturation of  N(y) for different non-linear equations
    (hard line - eq.(\ref{eik}, long-
   dashed line - eq. (\ref{u-matr}), and short-dashed - eq. (\ref{tangh})) }
\label{fig1}
\end{figure}

\section{Minimum saturation of the Soft plus Hard Pomeron model}

Let us take, as an  example, the Soft plus Hard pomeron  model
\cite{clms3,DoLa}
which includes two simple poles
to describe $pp$ and $\bar p p$ scattering at high $s$.
In this case, the $pp$-elastic scattering amplitude is
proportional to the hadron form factors and can be approximated
at small $t$ by:
\begin{eqnarray}
 T(s,t)\sim &&  [\ h_{1} \ (s/s_0)^{\epsilon_1}
           e^{\alpha^{\prime}_1 \  t \ \log (s/s_0)} \nonumber \\
   + h_{2} \ (s/s_0)^{\epsilon_2}
           &&    e^{\alpha^{\prime}_2 \  t \ \log (s/s_0)} ]
   \ F^2(t).   \label{ampl}
\end{eqnarray}
where $h_1=4.7$  and $h_2 = 0.005$ are the coupling of the soft
 and hard pomerons, and $\epsilon_1 =0.072$, $\alpha^{\prime}_1=0.25$
GeV$^{-2}$,
 and  $\epsilon_2=0.45$, $\alpha^{\prime}_2=0.20$ GeV$^{-2}$
 are the intercepts and the slopes of the two pomeron trajectories,
and  where $s$ contains implicitly the phase factor $\exp(-i \pi/2)$.
$F^2(t)$ is  the square of
 the Dirac elastic form factor.
It can be approximated by sum of three exponentials.  

We then obtain in the impact parameter representation
a specific form for the profile function $\Gamma(b,s)$.\cite{dif04}
One finds that for $\sqrt{s}\approx 1.5 \ $TeV and at small $b$,  $\Gamma(b,s)$
 reaches  the black disk limit.

Note that one cannot simply cut the profile function sharply
as this would lead
to a non-analytic amplitude, and to specific diffractive patterns
in the total cross section and in the slope of
the differential cross sections. Furthermore, we have to match at
 large impact parameter the behaviour of the unsaturated
profile function.
 We have to use some specific matching patters
which softly interpolate between both
regimes.
The interpolating function gives unity in the
large impact parameter region
and forces the profile function to approach the saturation scale $b_s$
as a Gaussian.

We show in Fig. 2 the possible behaviours of the total cross section
at very high energies, depending on $\epsilon =\alpha(0)-1$ and on
the unitarization scheme.
We also show there the result of a simple eikonalisation, where
we took
\begin{eqnarray}
     T_{pp}(s,t=0)
       \sim
  \int d^2b[1- e^{h_{eik} G(s,b)}]
\label{eik-pf}
\end{eqnarray}
and where $h_{eik} = 1.2$ was chosen so that the values
 $\sigma_{tot}$
determined  by the overlapping function (\ref{ampl}),
 and by eikonalisation (\ref{eik-pf})
 are equal at  $\sqrt{s}=50 \ $GeV.

\vskip -1.cm
\begin{figure}
\centerline{\psfig{file=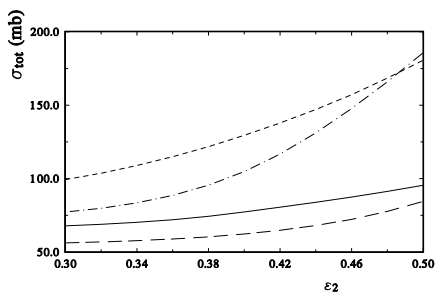,width=2.5in}}
\caption{The total cross section at $\sqrt(s) = 1.8 \ $TeV   and
  $\sqrt{s}=14 \ $TeV (lower two and upper two lines) for the
saturated amplitude (plain and short dashes) and
   for an eikonalized amplitude (long dashes and dash-dots). }
\label{fig2}
\end{figure}

\vskip -2.cm
\begin{figure}
\centerline{\psfig{file=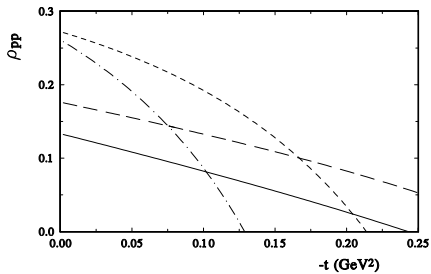,width=2.5in}}
\caption{The ratio of the real to the imaginary part of the amplitude as a function
 of $t$ for the saturated amplitudes at various energies: $100, \ 500,  \ $GeV, and
 $5, \ 14 \ $TeV  (plain, short-dashed, dot-dashed lines correspondingly  }
\label{fig3}
\end{figure}

To demonstrate the presence of saturation at the LHC,
one can also consider the slope of
the differential elastic cross section, which we show in Fig.3.
We see that saturation increases its value at small $t$ (this is in fact
unavoidable for any saturation scheme), and predicts a
fast drop around $|t|=0.25$ GeV$^2$, when one enters the dip region.

In conclusion, we have shown that the most usual unitarization schemes could be cast into
  differential equations which are reminiscent of the saturation equation
.\cite{balitsky,kovchegov}
  Such an approach can be used to build new unitarization schemes
  and may also shed some light on the physical processes responsible for
 the saturation regime.

\section*{Acknowledgments} O.V.S. acknowledges the support
of FRNS (Belgium) for visits
  to the University of Li\`ege where part of this work was done. 


\begin{thebibliography}{99}

 \bibitem{clms2}J.-R. Cudell, A. Lengyel, E. Martynov, O.V. Selyugin,
                    Phys. Lett. {\bf B587} (2004) 78--86.
\bibitem{clms3}   J.-R. Cudell, A. Lengyel, E. Martynov, O.V. Selyugin,
       Nucl. Phys. {\bf A755} (2005) 587-590.
        [arXiv: hep-ph/0501288].
\bibitem{clms1} E. Martynov, J.R. Cudell, O.V. Selyugin
                  Eur. Phys. J. {\bf C33} (2004) S533.
\bibitem{muller} A.H. Mueller, Nucl. Phys. {\bf B335} (1990) 115.

\bibitem{CBWwud} K. Golec-Biernat and M. Wusthoff, Phys. Rev.
    {\bf D60}  (1990) 114023.
\bibitem{umat} A.A. Logunov, V.I. Savrin, N.E. Tyurin, and O.A. Khrustalev,
  Theor. Mat. Fiz. {\bf 6} (1971) 157.

\bibitem{trosh} S.M. Troshin, and N.E. Tyurin, Phys. Lett. {\bf B316}
  (1993) 316.

\bibitem{dif04}
 J.~R.~Cudell and O.~V.~Selyugin,
 Czech. J. Phys. {\bf 54} (2004) A441-A444
[arXiv: hep-ph/0309194].

\bibitem{grib} L.V. Gribov, E.Levin and M. Ryskin, Phys. Rep.
  {\bf D49} (1994);
  A. Mueller and J.W. Qiu, Nucl. Phys. {\bf B286} (1986) 427.
\bibitem{mcler} L. McLerran and R. Venugopalan, Phys. Rev. {\bf D49} (11994)
 2233;  Phys. Rev. {\bf D49} (1994) 3352.


\bibitem{balitsky} J. Balitsky, Nucl. Phys. {\bf B 463} (1996) 99.

\bibitem{kovchegov} Y.V. Kovchegov, Phys. Rev. {\bf D60} (1999) 0340008 ;
  Phys. Rev. {\bf D61} (2000) 074018.

\bibitem{cs7}   J.-R. Cudell, O.V. Selyugin,
         Nucl. Phys. B (Proc. Suppl.) {\bf 146},  185-187 [arXiv: hep-ph/0412338].

\bibitem{DoLa}  A. Donnachie and P. Landshoff, Nucl. Phys.  {\bf B267}
 (1986) 690. 


\bibitem{cs8} J.-R. Cudell, O.V. Selyugin,
        Czech. J. Phys. {\bf 55}  (2005) A235-A242.

\bibitem{cs9}   J.-R. Cudell, A. Lengyel, E. Martynov, O.V. Selyugin,
        Nucl. Phys. B (Proc. Suppl.) {\bf 152} (2005)  182-187.




\end{thebibliography}
\end {document}